# High-Energy Spectral Complexity from Thermal Gradients in Black Hole Atmospheres

J. G. Skibo[1] and C. D. Dermer

E. O. Hulburt Center for Space Research, Code 7653, Naval Research Laboratory, Washington, DC 20375-5352

## ABSTRACT

We show that Compton scattering of soft photons with energies $\sim 100$ eV in thermally stratified black-hole accretion plasmas with temperatures in the range $\sim 100$ keV - 1 MeV can give rise to an X-ray spectral hardening above $\sim 10$ keV. This could produce the hardening observed in the X-ray spectra of black holes, which is generally attributed to reflection or partial covering of the incident continuum source by cold optically thick matter. In addition, we show that the presence of very hot ($kT \simeq 1$ MeV) cores in plasmas leads to spectra exibiting high energy tails similar to those observed from Galactic black-hole candidates.

*Subject headings:* Black Hole Physics — Galaxies: Seyfert — Radiation Mechanisms: Thermal

## 1. Introduction

*HEAO-1* and *EXOSAT* observations of the 2-10 keV X-ray spectra of Seyfert AGNs (see Mushotzky, Done, & Pounds 1993 for a recent review) had been well represented by single power laws with energy spectral index $\alpha \simeq 0.7$ (Rothschild et al. 1983; Mushotzky 1984; Turner & Pounds 1989). Later observations with the *Ginga* satellite revealed X-ray spectra which, in addition to an underlying power law continuum, display strong 6.4 keV line emission from neutral Fe, absorption edges at $\sim 7$-8 keV from ionized Fe, and a spectral hardening above about 10 keV (Pounds et al. 1990). Such features are also present in the *Ginga* spectra of the Galactic black hole candidate Cygnus X-1 (Done et al. 1992). If these additional features are included in the modeling and the spectral hardening is attributed to Compton reflection, then the average intrinsic power law index of Seyferts is found to be closer to $\alpha \simeq 0.9$ (Nandra & Pounds 1994). At gamma-ray energies, observations of Seyfert galaxies made with OSSE on the Compton Gamma Ray Observatory (CGRO) reveal spectra which generally display an exponential cut off or a spectral softening with an e-folding energy of $\sim 50$ keV - several hundred keV (Maisack et al.

---

[1] NAS/NRC Research Associate

1993; Johnson et al. 1994; Zdziarski et al. 1995). No $e^+$-$e^-$ annihilation features have been detected from these sources, indicating that the primary emission is most likely due to thermal Comptonization in very hot ($kT \sim 100$ keV) plasmas. Furthermore, Monte Carlo simulations of Comptonization in $e^+$-$e^-$ pair balanced plasmas reveal that the observed spectral indices and time variablity are in accord with thermal plasmas consisting of no pairs (Skibo et al. 1995).

The Fe spectral features and the hardening at 10-100 keV are generally interpreted as the reprocessing signatures of the primary power law continuum by cold optically thick matter (Lightman & White 1988; Guilbert & Rees 1988). This medium absorbs the X-ray emission and produces the 6.4 keV Fe K$\alpha$ florescence line, and through a combination of Compton downscattering, bound-free absorption, and the diminution of the Compton scattering cross section at higher energies, produces a "reflection hump" at 10 - 100 keV (George & Fabian 1991; Matt et al. 1991). One scenario is that the primary high-energy continuum is produced via Comptonization of soft photons from an accretion disk in a coronal plasma sandwiching the disk (Haardt & Maraschi 1991; 1993), with the reflecting medium being the disk itself. This picture appears to be supported by recent ASCA observations of MCG-6-30-15 (Tanaka et al. 1995), NGC 4151 (Yaqoob et al. 1995), NGC 5548 and IC 4329A (Mushotzky et al. 1995), where the 6.4 keV line displays an asymmetry which could be produced by relativistic effects in a black hole accretion disk. It is puzzling, however, as to why the inferred disk orientations are always near face on, especially in the case of NGC 4151 (Yaqoob et al. 1995). Other models place the reflecting medium farther out, for example, in an obscuring torus (Krolik, Madau, & Życki 1994; Ghisellini, Haardt, & Matt 1994) or in dense blobs surrounding the central source (Nandra & George 1994). An alternative explanation for the Fe features and 10 - 100 keV hardening is that the central source is partially covered by cold dense material which is optically thick below $\sim 6$ keV (Holt et al. 1980; Matsuoka et al. 1990; Piro et al. 1990).

A recent reanalysis of *HEAO-1* data reveals that strong Fe K$\alpha$ lines tend not to be associated with the presence of a 10 - 100 keV hardening (Weaver, Arnaud, & Mushotzky 1995). Furthermore, an unphysical amount of reflection (i.e., a reflector subtending $> 4\pi$ steradians) is inferred in some cases and attributed to time lags between the direct and reflected continuum. In the scenarios just described, the Fe K$\alpha$ line and the 10 - 100 keV hardening are produced in the same medium, so that the presence of the Fe line and the reflection feature should be correlated. Here we use a Monte Carlo simulation to show that the observed X-ray spectral hardening above $\sim 10$ keV can result entirely from Comptonization in the central black hole atmosphere and need not be attributed to reflection or partial covering of the central source of continuum emission. This effect occurs if soft photons with energies $\sim 100$ eV are injected into the Comptonizing region, and is enhanced if radial thermal gradients are present in the plasma. Furthermore, we show that if the plasma contains a very hot ($kT \simeq 1$ MeV) inner core, then a high energy excess above $\sim 1$ MeV is formed. Such features have been observed in the soft gamma ray spectra of the Galactic black hole candidates Cygnus X-1 (McConnell et al. 1994) and GRO J0422+32 (Roques et al. 1994; van Dijk et al. 1995). Our model can be tested with correlated observations of hard X-ray, soft gamma-ray, and Fe line emission. A spectral hardening near 10 keV which correlates with the high-energy excess yet varies in an uncorrelated fashion with the Fe K$\alpha$ emission would give evidence for the proposed scenario.

## 2. Results

The picture we consider is illustrated in Figure 1. A geometrically thin, optically thick accretion disk (Shakura & Sunyaev 1973) forms around a central black hole and, due to instabilities in the hot inner regions, swells into a geometrically thick, optically thin, gas-pressure dominated hot plasma (Shapiro, Lightman & Eardley 1976). Just as the standard disk solutions exhibit temperatures that increase with decreasing radius, we likewise argue that temperature gradients will be present in the inner region. Deeper in the gravitational potential well the protons are hotter and, therefore, impart more energy to the electrons through Coulomb heating. A major source of electron cooling is Compton scattering of the soft disk radiation. As these photons penetrate the thermal plasma, their mean energy rises as they extract energy from the electrons. Hence the outer regions of the Comptonizing plasma are more efficiently cooled, resulting in a negative radial temperature gradient.

We have constructed a Monte Carlo simulation model to examine the effects that temperature gradients can have on the emergent radiation spectrum. The details of the thermal Comptonization code are given by Skibo et al. (1995). For simplicity, we considered a two-zone spherical geometry with outer radius $R_{out}$ and total Thomson depth $\tau_T$ along the radial direction (see Figure 1). The sphere is divided into an outer zone ($R_{in} \leq r < R_{out}$) with temperature $\Theta_{out} \equiv kT_{out}/m_e c^2 = 0.2$ ($kT_{out} \simeq 100$ keV) and a hotter inner zone ($R_S \ll r \leq R_{in}$) with temperature $\Theta_{in} = 0.6$ ($kT_{in} \simeq 300$ keV), and we set $\tau_T = 1$. Here we neglect general relativistic effects on spectral formation in the region near the Schwarzschild radius $R_S$. External soft photons having a blackbody energy distribution with temperature $kT_{bb} = 100$ eV were injected radially inward at the surface of the sphere. The values of $\Theta_{in}$, $\Theta_{out}$ and $\tau_T$ were chosen to yield X-ray energy spectral indicies near unity. Harder spectra (i.e. $\alpha \simeq 0.5$), such as that seen in several sources (Turner & Pounds 1989), can easily be produced by increasing the optical depth or temperature.

In Figure 2 we show the calculated spectra with $R_{in}/R_{out} \ll 1$ and $R_{in}/R_{out} = 1/2$. We plot the quantity $\epsilon^2 \, dN/d\epsilon$, which is a measure of the power per logarithmic interval of energy. The normalization is such that the integral number of photons is unity. The thick solid curves are the emergent Compton-scattered photon spectra. The dotted lines represent the X-ray power law fits to the Monte Carlo results in the 2-10 keV energy range. The thin solid curves are the individual scattering spectral profiles. Those at the far left of each panel correspond to photons escaping with zero scatterings, the next to the right, single scattering, etc., up to the seventh scatter (the thick solid curves include all orders of scattering). Fig. 2a displays the case of constant temperature $\Theta = 0.2$ (i.e. $R_{in}/R_{out} \ll 1$). A weak spectral hardening from the formation of the Wien peak is visible in this case. It is not possible, however, to produce a strong spectral hardening between $\sim 10$ - $100$ keV (implying $\Theta \simeq 0.2$) from Wien formation in a single component plasma with $\alpha \simeq 1$, because $\tau_T$ must be $\simeq 1$ in order to satisfy these conditions.

As can be seen from Fig. 2b, the presence of the hot inner core causes the 2-10 keV power-law to harden and the excess above 10 keV to become more pronounced. From the way in which the individual scattering profiles constitute the total spectrum it can be seen that the spectral hardening above $\sim 10$ keV results from asymmetries in the profiles corresponding to low number of scatterings. That is, as $R_{in}/R_{out}$ increases, the profiles corresponding to two or three scatterings are skewed towards higher energies. This results from the occasional scattering of photons with electrons in the hot core. Hence, the profiles for low number of scatterings form a sort of pedestal over which the saturating profiles for higher number of scatterings lie. This effect clearly depends on the photon injection, which we take here to have a blackbody spectrum with $kT_{bb} = 100$ eV. We repeated the calculation with $kT_{bb} = 10$ eV and found that spectral hardening was less pronounced and ocurred approximately a decade lower in energy than the case where $kT_{bb} = 100$ eV.

In Figure 3 we show the spectrum resulting from a plasma with a very hot and compact inner zone. That is, we set $kT_{bb} = 100$ eV, $\Theta_{out} = 0.2$ ($kT_{out} \simeq 100$ keV), $R_{in}/R_{out} = 0.2$ and $\Theta_{in} = 2$ ($kT_{in} \simeq 1$ MeV). It can be seen that the superposition of individual scattering profiles produce a secondary peak at higher energy due to photons scattering in the hot core. The saturation of these secondary peaks result in a high energy tail. Because the low optical depth of the hot inner region admits pair-free solutions (Skibo et al. 1995), we are justified in neglecting pairs in our simulations, provided we consider sufficiently low luminosity plasmas.

The plasma parameters in Figure 2b were chosen to show clearly the effect of the hard X-ray spectral hardening. The peak in luminosity is seen to occur at energies of about 200-300 keV in this case. This energy is higher than that implied by the simultaneous ROSAT and OSSE observations (Madejski et al. 1995) or the average Seyfert spectrum (Zdziarski et al. 1995). The location of the luminosity peak can be lowered without reducing the magnitude of the $\sim 10$ keV hardening by, for example, decreasing the inner core temperature, increasing its volume and adjusting the optical depth. The inclusion of general relativistic effects would also tend to redshift the emission peak, although these same effects would also lower the energy where the X-ray spectrum hardens.

## 3. Discussion

The mechanism introduced here for producing the X-ray spectral hardening arises entirely from the process of Comptonization in the central black hole atmosphere and does not require a reflection component. However, the presence of reprocessing material is indicated by atomic emission and absorption features, most notably line emission at 6.4 keV from neutral Fe. In the Ginga Seyfert data, for instance, the 6.4 keV line emission is observed to have equivalent widths in the range 100-150 eV, requiring neutral hydrogen columns of $\sim 10^{23}$ cm$^{-2}$, assuming solar Fe abundance (Nandra & Pounds 1994). However, equivalent widths as high as 300 eV are inferred from the recent ASCA data when the shape of the line in taken into account (Fabian et al. 1995). The gas reponsible for the line emission cannot totally cover the source due to the lack of absorption at soft X-ray energies which generally limits the neutral column to $\lesssim 10^{21}$ cm$^{-2}$ (Nandra & Pounds 1994). The exception is NGC 4151 where the neutral gas columns implied by the low energy absortion and Fe K$\alpha$ florescence are consistent if Fe is overabundant by a factor $\sim 2$ (Yaqoob & Warwick 1991). The X-ray spectra of some Seyferts also display an absorption edge near $\sim 7 - 8$ keV due to ionized gas in the line of sight ("warm absorber") with inferred columns of $\gtrsim 10^{23}$ cm$^{-2}$ (Nandra & Pounds 1994). However, the inferred columns from the Fe K$\alpha$ edge depend strongly on the underlying continuum. A more reliable estimate is obtained from measurements of the low energy absorption which imply columns of $10^{21-23}$ cm$^{-2}$ in about half the Seyferts observed with EXOSAT (Turner & Pounds 1989). In addition, some narrow emission line galaxies show absorbing columns of $\gtrsim 10^{23}$ cm$^{-2}$ (Warwick et al 1993).

The neutral columns ($\gtrsim 10^{23}$ cm$^{-2}$) required to produce the 6.4 keV emission, together with the lack of soft X-ray absorption, could result from optically thick ($\gtrsim 10^{24}$ cm$^{-2}$) reprocessing material in a flattened accretion-disk geometry or in the form of blobs which partially cover the central source (Nandra & George 1994). But the implied correlation between the strength of the Fe line and the reflection component has not been demonstrated and, moreover, an unphysical solid angle for reflection is required in many cases (Weaver et al. 1995). We have therefore proposed as an alternative to the reflection scenario that the 10 - 100 keV spectral hardening is the result of temperature gradients in the black hole atmosphere. As shown by our simulation in Fig. 2a of a single temperature plasma irradiated by UV soft photons, X-ray spectral complexity is unavoidable in any case. The required equivalent widths of the 6.4 keV emission line, as well as the absorption edge near $\sim 7 - 8$ keV, could be produced in our scenario by an optically thin circumnuclear environment consisting of neutral clouds with total column $\lesssim 10^{24}$ cm$^{-2}$ which are embedded in an ionized medium. If the clouds subtend a solid angle fraction of $\sim 50\%$ as seen from the source and have modestly enhanced Fe abundance (few times solar), then Fe features can be explained without introducing an absorption edge at 7.1 keV or Fe L-shell absorption at soft X-ray energies. A flattened cloud distribution, showing increased absorption with greater inclination angles, would be consistent with the unification of Seyfert 1 and 2 AGNs by orientation. We point out that in the reflection scenario extra-solar abundances are also occasionally invoked (Tanaka et al. 1995; Yaqoob et al. 1995).

The presence of an accretion disk should produce a reflection component and Fe line emission at some level. Thus both thermal Comptonization and reflection could be responsible for the $\sim 10$ keV spectral hardening, which would decrease the implied reflection solid angles. Furthermore, if the intrinsic continuum already has a hard X-ray excess in the intrinsic spectrum from thermal gradients in the Comptonizing plasma, then larger Fe K$\alpha$ equivalent widths will result from the increased number of photons above the Fe K$\alpha$ absorption edge. Using the results of George & Fabian (1991), we estimate that this effect will add $\simeq 30$ eV to the equivalent width if the disk is viewed nearly face-on. Any Fe K line emission from the disk, in addition to that from other matter, will reduce the required Fe abundance, possibly to solar. On the other hand, X-ray observations of high luminosity quasars (Williams et al. 1992; Nandra et al. 1995; Turner et al. 1995) as well as NGC 4151 (Maisack & Yaqoob 1991) do not show a $\sim 10$ keV spectral hardening, which implies that a reflection component is not always present.

The reflection or Comptonization mechanisms for producing the observed X-ray spectral hardening can be distinguished by correlated variability measurements of the Fe K line and the hard X-ray excess. In the reflection scenario, variations in the Fe K line could lag variations of the intrinsic source continuum (e.g., Matt & Perola 1992), but temporal variations in the Fe K line and the $\sim 10$ keV excess should vary synchronously: significant variations between the light curves of these features would be difficult to reconcile. On the other hand, if the X-ray spectral hardening results from Comptonization and the Fe features from reprocessing, then much more general behavior of the light curves is possible and the two features could vary in an entirely uncorrelated manner. The time scale of the temporal lag between the Fe K line and the $\sim 10$ keV excess would set an upper limit on the separation between the central continuum source and the reprocessing material. No lag would place the reprocessing material very close to the central source (although this behavior is not distinguishable from the reflection interpretation). In addition, the two models could be distinguished by the high-energy spectral shapes. For the reflection scenario, the peak luminosity is at a well-defined energy given the intrinsic source spectra, whereas the peak luminosity in the Comptonization scenario depends on the properties of the Comptonizing plasma.

Whereas reflection always produces a spectral hardening in the 10-100 keV range, the Comptonization scenario requires that the soft photons must be injected at energies on the order of a few hundred eV (i. e., $kT_{bb} \simeq 100$ eV) to produce the hardening in the 10-100 keV range. The soft excesses seen in about 30% of hard X-ray selected AGNs (Turner & Pounds 1989) can be modeled by a blackbody with $kT_{bb} \lesssim 150$ eV (Urry et al. 1989), so these observations are consistent with the soft photon injection energy used here. As mentioned above, a lower injection energy results in a spectral hardening at lower energies of diminished maginitude. For injection energies below about 20 eV, thermal Comptonization produces the spectral hardening $\lesssim 2$ keV. The simultaneous observation of a strong X-ray spectral hardening and low soft photon injection energies would therefore be difficult to reconcile in the thermal Comptonization scenario. But the weak spectral hardening associated with Wien peak formation (compare Fig. 2a) depends only on the Comptonizing plasma temperature and not on the soft photon injection energy.

Because Comptonization is capable of producing the spectral hardening at other energy ranges, depending upon the mean energy of the incident soft photons, it can explain other features seen in black hole spectra. High energy tails ($\gtrsim 1$ MeV) have been reported from the Galactic black hole candidates Cyg X-1 (McConnell et al. 1994) and GRO J0422+32 (Roques et al. 1994; van Dijk et al. 1995). We have shown that the presence of a hot central core in the Comptonization plasma gives rise to such high energy tails. In another interpretation these putative features are attributed to a spectral component arising from proton-proton $\pi^0$ production in addition to thermal Comptonization (Jourdain & Roques 1994).

To summarize, we have shown that Comptonization in plasmas containing thermal gradients can account for the 10 - 100 keV spectral hardening observed from Seyfert AGNs and the hard tails observed from Galactic black hole candidates. Spectral complexity is produced even by thermal Comptonization from uniform temperature plasmas, so that the use of an underlying power-law approximation for spectral modelling is not generally warranted. This scenario can be tested by observations of Fe K line features and $\sim 10$ keV and 100 keV-1 MeV emission. If the spectral hardening does arise from the appearance of a thermal gradient in the central hot plasma, then the magnitude of the high energy excess between $\sim 100$ keV and 1 MeV should correlate with the $\sim 10$ keV excess. Correlated observations with XTE, SAX, and the OSSE telescope on the *Compton Observatory* could test these ideas for the brightest Seyfert galaxies such as IC 4329A and NGC 4151 as well as the galactic black hole candidates Cyg X-1 and GRO J0422+32.

We are grateful to Dr. Tahir Yaqoob for providing useful suggestions and detailed criticism.

Fig. 1.— Schematic illustration of an accreting black hole. Due to instabilities in the disk the central region swells into a geometrically thick hot thermal plasma which Comptonizes soft photons from the accretion disk to higher energies. In the Monte Carlo Comptonization simulation the central region is assumed to have spherical geometry with two radial zones at different temperatures.

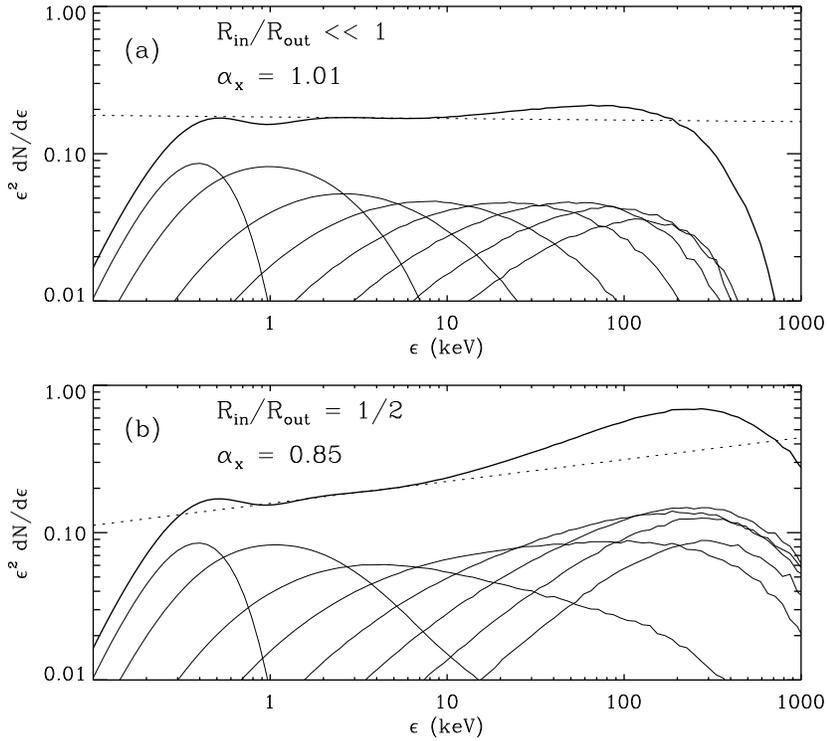

Fig. 2.— Spectra of emergent Comptonized photons (solid curves) and individual scattering spectral profiles (thin solid curves). The dotted lines represent the X-ray power law fits to the Monte Carlo spectra in the 2-10 keV energy range. The 2-10 keV X-ray energy spectral indices are given nn the figure legends. In (a) are shown spectra for the case of constant temperature ($\Theta = 0.2$ i.e. $R_{in}/R_{out} \ll 1$). Spectra in (b) show the effects of a hot ($\Theta = 0.6$) core of size $R_{in}/R_{out} = 1/2$. Note the hardening above the 2-10 keV power-law extensions in both cases.

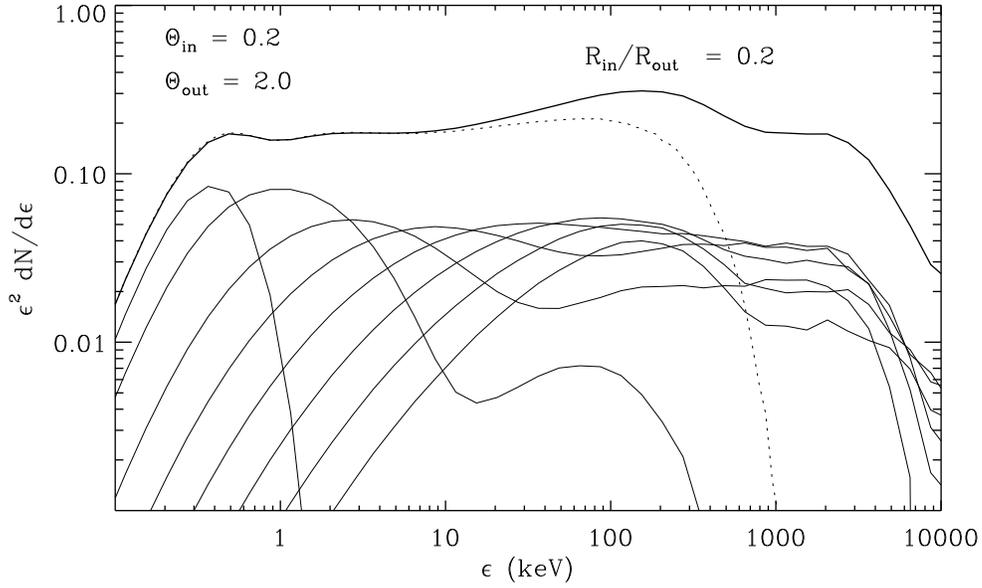

Fig. 3.— Calculated emergent spectrum from a Comptonization plasma with a very hot and compact core. The parameters in this simulation are $kT_{bb} = 100$ eV, $R_{in}/R_{out} = 0.2$, $\Theta_{in} = 2$, and $\Theta_{out} = 0.2$. The solid curves have the same meaning as in Figure 2. The dotted curve represents the emergent spectrum calculated in the absence of the hot core (i.e. $R_{in}/R_{out} \ll 1$).